\documentclass[11pt]{article}
\usepackage{float,amssymb}
\usepackage{graphicx}% Include figure files
\newcommand{\beq}{\begin{equation}}
\newcommand{\eeq}{\end{equation}}
\newcommand{\beqa}{\begin{eqnarray}}
\newcommand{\eeqa}{\end{eqnarray}}

\newcommand{\vs}{\vspace{-0.20cm}}
\begin{document}

%%\noindent Accepted for publication in Phys. Lett. {\bf B} 

\begin{flushright}
{\tiny  HISKP-TH-04/04} \\
{\tiny  UK/TP-2004-05} \\
\end{flushright}

\vspace{.6in}

\begin{center}

\bigskip

{{\Large\bf Radiative neutron  \boldmath{$\beta$}--decay in effective 
          field theory
    }}

\end{center}

\vspace{.3in}

\begin{center}
{\large 
V\'eronique Bernard$^\star$\footnote{email: bernard@lpt6.u-strasbg.fr},
Susan Gardner$^\dagger$\footnote{email: gardner@pa.uky.edu},
Ulf-G. 
Mei{\ss}ner$^\ddagger$$^\ast$\footnote{email: meissner@itkp.uni-bonn.de},
Chi Zhang$^\dagger$\footnote{Present Address: 
124 Sherman Hall, The Department of Physiology and Biophysics, 
SUNY, Buffalo, NY 14226. 
%email: chi@pa.uky.edu
}
}

\vspace{1cm}

$^\star${\it Universit\'e Louis Pasteur, Laboratoire de Physique
            Th\'eorique\\ 3-5, rue de l'Universit\'e,
            F--67084 Strasbourg, France}

\bigskip

$^\dagger${\it Department of Physics and Astronomy, University of Kentucky\\
Lexington, Kentucky 40506-0055, USA}\\

\bigskip

$^\ddagger${\it Universit\"at Bonn,
Helmholtz--Institut f\"ur Strahlen-- und Kernphysik (Theorie)\\
Nu{\ss}allee 14-16,
D-53115 Bonn, Germany}

\bigskip

$^\ast${\it Forschungszentrum J\"ulich, Institut f\"ur Kernphysik 
(Theorie)\\ D-52425 J\"ulich, Germany}

\bigskip

\bigskip

\end{center}

\vspace{.4in}

\thispagestyle{empty} 

\begin{abstract}\noindent 
We consider radiative $\beta$--decay of the neutron in heavy baryon
chiral perturbation theory, with an extension including explicit $\Delta$
degrees of freedom. We compute the photon energy spectrum as well as
the photon polarization; both observables are 
dominated by the electron bremsstrahlung contribution. 
Nucleon-structure effects not encoded in the weak coupling constants
$g_A$ and $g_V$ are determined at next-to-leading order in the
chiral expansion, and enter at the ${\cal O}(0.5\%)$-level, making 
a sensitive test of the Dirac structure of the weak currents
possible. 
\end{abstract}

\vfill

\pagebreak

\noindent {\bf 1.}  Experimental studies of $\beta$--decay at low
energies have played a crucial role in the rise 
of the Standard Model (SM)~\cite{review}. In recent years, 
continuing, 
precision studies of  neutron $\beta$--decay have 
been performed, to better both the determination of the neutron
lifetime and of the correlation coefficients. 
Taken in concert, 
these measurements yield the weak coupling constants
$g_V$ and $g_A$\footnote{Precise
definitions of the various form factors and couplings follow.};
$g_V$, in turn, yields the Cabibbo-Kobayashi-Maskawa (CKM) matrix element
$V_{ud}$ and, with the empirical values of $V_{us}$ and $V_{ub}$, 
the most precise test of the unitarity of the CKM matrix. 
As the neutron measurements improve, further SM tests become possible, such 
as a precision test of the CVC hypothesis, 
as well as of the absence of second-class currents, 
yielding, generally, improved constraints
on the appearance of non-$V-A$ currents~\cite{chisvg}. 

To realize a SM test to a precision of $\sim 1\%$ or better 
requires the 
application of radiative corrections~\cite{sirlin}. 
For example, a new measurement of the $A$ correlation coefficient in neutron
$\beta$--decay, with the world average values for the 
neutron lifetime, $V_{us}$, and $V_{ub}$~\cite{pdg2000}, 
yields $1 - \sum_{j=d,s,b} |V_{uj}|^2= 0.0083(28)$~\cite{abele}, 
indicating
a deviation of $3\sigma$ from CKM unitarity. 
The significance of the deviation from unitarity depends 
on the radiative corrections and their surety. 
One component of 
such, the ``outer'' radiative correction, 
is captured by electromagnetic 
interactions with the charged, 
final-state particles, in the limit in which their structure is
neglected. In this, neutron radiative $\beta$--decay enters, and
we consider it explicitly. 
We find neutron radiative
$\beta$--decay 
interesting in its own right, though the 
process has yet to be
observed --- only an upper bound exists~\cite{mbeck}. 
Anticipating its measurement, however, 
and as the precision of such improves, 
we can (i) hope to effect an 
alternative  determination of the weak couplings $g_V$ and $g_A$. 
The photon energy spectrum in neutron radiative $\beta$--decay in leading order
is characterized by contributions proportional to  $g_V^2 + 3 g_A^2$
and to $g_V^2 - g_A^2$, so that $g_V$ and $g_A$ can be determined, 
though $(g_V^2 - g_A^2)/(g_V^2 + 3g_A^2) \sim 0.10$. 
(ii) We can study the hadron matrix elements 
in subleading order, ${\mathcal O}(1/M)$, with $M$ the 
neutron mass. Here, we
note the connection to radiative muon capture on the proton, which 
permits the determination of the 
induced pseudoscalar coupling constant $g_P$. The only
measurement thus far 
of radiative muon capture \cite{TRIUMF} yields a result for $g_P$
which is significantly at odds with the 
chiral perturbation theory 
prediction \cite{BKMgp}.  
For recent reviews containing extensive discussions of possible resolutions
to this problem, see Refs.~\cite{BEM,GF}. The same hadronic matrix elements,
calculated in Ref.~\cite{BHM} 
in the framework of an effective field theory (EFT) of
nucleons, pions, and external sources (and $\Delta$s), appear
in radiative neutron capture, albeit at much smaller momentum transfers.
Consequently, one could 
integrate out the $\Delta$s and even the 
pions from the EFT, resulting in an equally precise calculational
tool but with no direct access to and thus test of 
the chiral structure of QCD at low
energies.\footnote{Alternatively, one could  use a nonrelativistic EFT for the
calculation and then perform matching to the amplitudes evaluated in 
heavy baryon chiral perturbation or in the small scale expansion. 
We prefer, however, to work  with an EFT including explicit 
pions (and $\Delta$s).} 
(iii) We can use neutron radiative $\beta$--decay to test the Dirac
structure of the weak current, through the determination of the 
circular polarization
of the associated photon~\cite{gaponov}. 
As recognized shortly after the discovery
of parity violation in $\beta$--decay~\cite{wu}, the photon emitted
in associated radiative processes 
should be circularly 
polarized~\cite{cutkosky,marglau}. 
In integrating over the phase space, it becomes apparent
that the photon becomes $\sim$100\% polarized only when its energy grows 
large; in our explicit calculations we confirm
that the predictions of Ref.~\cite{marglau} for internal bremsstrahlung,
i.e., for radiative orbital electron capture of $S$-state electrons, 
are germane to radiative $\beta$--decay as well. 
This prediction follows 
from a perfectly 
right-handed anti-neutrino and from 
the absence of scalar, tensor, and pseudoscalar 
interactions in leading order. 

In this letter,  we perform a systematic analysis of neutron radiative
$\beta$--decay in the framework of heavy baryon chiral perturbation theory
(HBCHPT) \cite{JM,BKKM,BKMrev} and in the small scale 
expansion (SSE) \cite{HHK}, 
including all terms in $O(1/M)$,
i.e., at next--to--leading order (NLO) in the small parameter $\epsilon$.
Here, $\epsilon$ collects all the 
small external momenta and quark (pion) masses, 
relative to the heavy baryon mass $M$, which appear 
when HBCHPT is utilized; in case of the SSE, 
such is supplemented by the 
$\Delta(1232)$--nucleon mass splitting, 
relative 
to the nucleon mass, as well.  
These systematic EFTs allows one
to calculate the recoil-order corrections 
in a controlled way. 
In order to assess the size of the recoil-order corrections, 
we compare with the pioneering work 
of Ref.~\cite{gaponov}, in which such effects have been neglected.
In that calculation, the standard parameterization of the 
hadronic weak current in terms of the weak coupling constants 
suffices to capture the hadron physics. 
No reference to photon emission from the effective four--fermion vertex 
is found in these papers. 
Here, we include all terms in $O(1/M)$, utilizing the 
framework of HBCHPT and the SSE for the actual calculations. In fact, 
the pertinent two-- and four--point functions can be taken directly 
from Ref.~\cite{BHM}, after relabelling the momenta and such.
\medskip

\begin{figure}[ht]
\begin{center}
\includegraphics[height=1in]{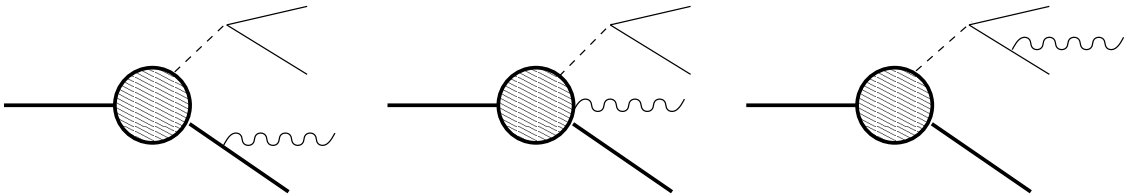}
\vspace{-0.2cm}
\caption{Contributions to $n\to p e^- \bar \nu_e \gamma$ in 
${\cal O}(G_F)$; 
heavy lines denote nucleons, light lines denote leptons, wiggly
lines denote photons, and the
shaded circle denotes the effective weak vertex. 
\label{fig:brems}}
\end{center}
\end{figure}

\noindent {\bf 2.} 
First, we collect some definitions for the process under consideration,
\beq
n(p) \to p(p^\prime) + e^- (l_e) + \bar{\nu}_e (l_\nu) + \gamma (k)~,
\eeq
where $p, \,p^\prime, \,l_e, \,l_\nu,$ and $k$ denote 
the four--momentum of the neutron, proton, electron, anti-neutrino, 
and photon, 
respectively --- we denote the photon energy by $\omega$. 
In the
static approximation for the $W^-$-boson, which is appropriate here, the 
matrix element for radiative neutron $\beta$--decay 
decomposes into two pieces, 
\begin{eqnarray}
{\cal M} (n \to p e^- \bar\nu_e \gamma )&=&
                \langle\bar\nu_e\, e^- |J_\mu^-|0\rangle
                \,i\,\frac{g^{\mu\nu}}{M_W^2}
                \langle p|{\cal T}\, (V\cdot\epsilon^\ast V_\nu^+
                - V\cdot\epsilon^\ast A_\nu^+) |n\rangle
                \nonumber \\
             & &+\langle \bar\nu_e\, e^- \,\gamma|J_\mu^-| 0 \rangle
                \,i\,\frac{g^{\mu\nu}}{M_W^2} \langle p|V_\nu^+
                - A_\nu^+|n\rangle~, \label{me}
\end{eqnarray}
in terms 
of the leptonic 
weak current ($J^-$), as well as 
the hadronic vector ($V$) and axial--vector ($A$) currents;  
$\epsilon_\mu$ is the photon polarization vector. For later use, we
introduce the Fermi constant $G_F$ via $G_F={g_2^2\sqrt{2}}/{(8 M_W^2)}$,
where $M_W$ is the W--boson mass and $g_2$ is the usual 
SU(2)$_L$ gauge coupling constant. The first term in Eq.~(\ref{me}) 
includes bremsstrahlung from 
the proton, as well as radiation from the effective weak vertex, 
which includes radiation from the pion in flight, 
whereas the second 
term corresponds to bremsstrahlung from the electron in the final state.
These contributions are illustrated in Fig.~\ref{fig:brems}. 

\vspace{0.5cm}

We now discuss the leptonic and hadronic matrix elements appearing in
Eq.~(\ref{me}). The pertinent leptonic current matrix elements are 
\begin{eqnarray}
\langle \bar \nu_e \, e^- |J_\mu^-\,|0\rangle&=& - i\,
                             \frac{g_2}{\sqrt 8}\,
\bar{u}_{e}(l_e)\gamma_\mu ( 1 - \gamma_5) v_\nu(l_\nu) \,,
\label{lepcurr}  \\
\langle \bar \nu_e \, e^- \, \gamma |J_\mu^-\,|0\rangle&=& i\,
                             \frac{g_2 e}{\sqrt 8}\,
\bar{u}_{e}(l_e) \left( \frac{2\epsilon^\ast\cdot l_e 
- \slash{\!\!\!\!\!\;k}\,\, \slash{\!\!\!\!\!\;\epsilon^\ast}}{2 l_e \cdot k} 
\right) \gamma_\mu ( 1 - \gamma_5) v_\nu(l_\nu) \;, 
\label{lepcurre}  
\end{eqnarray}
whereas the most general form of the hadronic 
weak current matrix elements, consistent with the $V-A$ structure of 
the SM, is~\cite{GT58}
\begin{eqnarray}
\langle p\,|V_\mu^+\,|n\rangle&=&-i\,
                             \frac{g_2 }{\sqrt 8}\,\bar{u}_p(p^\prime)
                             \left[F_1(q^2)
\gamma_\mu - \frac {i}{2M}F_2(q^2) \sigma_{\mu \nu} q^\nu
+ \frac{F_3(q^2)}{2M}q_\mu 
\right]u_n(p)~, \label{Vcorr}  \\
\langle p\,|A_\mu^+\,|n\rangle&=&-i\,
                             \frac{g_2 }{\sqrt 8}\,\bar{u}_p(p^\prime)
                             \left[G_1 (q^2)\gamma_\mu\gamma_5
- \frac {i}{2M}G_2(q^2) \sigma_{\mu \nu} \gamma_5 q^\nu
- \frac{G_3(q^2)}{2M}q_\mu \gamma_5
\right]u_n(p)~, \label{Acorr} 
\end{eqnarray}
with $\sigma^{\mu\nu} = i[\gamma^\mu,\gamma^\nu]/2$ and 
 $q_\mu\equiv (p - p^\prime)_\mu$. 
We note that Eqs.(\ref{lepcurr}-\ref{Acorr})
employ conventional spinors, satisfying, e.g., 
$\sum_s u_e(l,s) {\bar u}_e(l,s) = \slash{\!\!\!\!\;l} + m_e$. 
The weak coupling constants $g_V$ and $g_A$, which appear
in leading order, are
defined via $F_1(0)\equiv g_V$ and $G_1(0)\equiv g_A$.
We note that 
$g_A/g_V\equiv \lambda =1.2670 \pm 0.0030$ 
as determined from the $A$ correlation coefficient 
in neutron $\beta$--decay~\cite{pdg2002}. 
In the SM, under an assumption of isospin symmetry, the CVC hypothesis 
relates the weak vector form factors to the (electromagnetic) 
Dirac and Pauli form factors; we recall that
the Dirac form factor is unity at $q^2=0$, so that 
$g_V \equiv (1+ \Delta_R^V)^{1/2}V_{ud}$, where 
$\Delta_R^V$ is a small, radiative correction~\cite{sirlin} and 
$V_{ud}$ is a Cabibbo-Kobayashi-Maskawa (CKM) matrix element. 
Moreover, the CVC hypothesis and isospin symmetry determines 
the weak magnetism term, namely, that 
$F_2(0)/F_1(0)=\kappa_v$, where $\kappa_v =3.706$ is 
the isovector nucleon anomalous magnetic moment; we have
neglected the possibility of an additional radiative correction
which is not common to $F_1$ and $F_2$. 
The second-class current
contributions $F_3(q^2)$ and $G_2(q^2)$ vanish at $q^2=0$ in this
limit, so that henceforth we omit any discussion of them entirely. 
Isospin is an approximate symmetry of the SM, so that corrections
to these expectations, save that of $F_1(0)$, are of
${\cal O}(R)$, 
where $R\approx (M - M^\prime)/{M_N}$, noting that $M$ and
$M^\prime$ are the neutron and proton mass, respectively, with 
$M_N \equiv(M + M^\prime)/2$ 
the average neutron-proton mass. Such corrections, however, 
are systematically of higher order in our power counting scheme 
and thus can be neglected to the order, ${\cal O}(1/M^2)$, in which we work. 
Usually the non--relativistic reduction of Eqs.(\ref{Vcorr},\ref{Acorr})
is done in 
the Breit frame. Here we give the non--relativistic 
strong matrix elements in the rest frame of the neutron where our
calculation is done: 
\begin{eqnarray}
\langle p|\,V_\mu^+\,|n\rangle \! &= \! &-i\,
\frac{g_2 
}{\sqrt 8}\ {\cal N}^\prime \,
\bar{p}_v(p^\prime) \left\{
\left(\frac{2 M}{E^\prime +M}F_1(q^2)
      -\frac{E^\prime -M}{E^\prime +M}F_2(q^2)\right)
v_\mu \right.\nonumber\\
& &-\left[\frac{1}{E^\prime +M}\left(F_1(q^2)+F_2(q^2)\right)
      -\frac{1}{2 M}F_2(q^2)\right]q_\mu\nonumber\\
& &\left.-\frac{2}{E^\prime +M}\left[S_\mu,S\cdot q\right]
      \left(F_1(q^2)+F_2(q^2)\right)
\right\} n_v(0)~, \label{pVn}\\
\langle p|\,A_\mu^+\,|n\rangle \! &= \!& -i\,
\frac{g_2 
}{\sqrt 8}\ {\cal N}^\prime \,
\bar{p}_v(p^\prime )\left\{G_1 (q^2)\left[2\,S_\mu + \frac{2\,S\cdot q\, v_\mu}
                  {E^\prime +M}\right] 
+ G_3(q^2)\,\frac{S\cdot q\,q_\mu}{M\left(E^\prime + M\right)}
\right\}n_v(0), \label{pAn}
\end{eqnarray}
where we expand 
Eqs.~(\ref{pVn},\ref{pAn}) to ${\cal O}(1/M^2)$ in all applications. 
Note that 
${\cal N}^\prime$ is the usual normalization factor of the proton wave
function, ${\cal N}^\prime =\sqrt{ {(E^\prime +M)}/{2M}}$ 
and $E^\prime$ 
is the proton energy. We have 
employed non-relativistic nucleon spinors, with normalization
$\sum_\sigma n_v(r,\sigma) {\bar n}_v(r,\sigma) = 
P_v^+(1 + v\cdot r/(2M))$, where $P_v^+\equiv (1 + \slash{\!\!\!\!\!\; v})/2$. 
We make use of the fact that in HBCHPT, and in the SSE, 
the nucleon four--momentum $p_\mu$ is written as $p_\mu = M v_\mu + r_\mu$, 
with $v_\mu$ the fixed four--velocity, subject to the constraint $v^2 = 1$ 
and $r \cdot v \ll M$. Furthermore, $S_\mu$ is the nucleon's (Pauli-Lubanski)
spin vector with $v \cdot S = 0$.
The explicit form of the four form factors appearing in the
above equations, expanded to next-to-leading order in HBCHPT and
in the SSE, 
can be taken from Refs.\cite{BKKM,BFHM,Fearing:1997dp} for HBCHPT and
from Ref.\cite{BFHM} in the SSE.
At the small momentum transfers of current interest, however, 
it suffices to replace the form factors with their values at 
zero $q^2$, though we do employ 
$G_3(q^2)/M=4 g_{\pi NN} F_\pi/(m_\pi^2 - q^2) - 2 \lambda M r_A^2/3$,
where the radiative corrections implicit to the use of $\lambda$
in this case are without numerical consequence. 
For reference, the induced pseudoscalar coupling constant,
$g_P$, is $g_P \equiv G_3 (-0.88 m_\mu^2)/2M$ with $m_\mu$ the
muon mass. 
We now turn to the vector--vector (VV) and vector--axial (VA) 
correlators, which we need to ${\cal O}(p^2)$ in 
HBCHPT, or to ${\cal O}(\epsilon^2)$ in the
SSE. Working in the  Coulomb gauge $\epsilon^\ast\cdot v= 0$ for the photon
and making use of the transversality condition $\epsilon^\ast\cdot
k=0$, we find 
\begin{eqnarray}
\langle p|\,{\cal T} \,V\cdot\epsilon^\ast V_\mu^+\,|n\rangle^{(2)}
\!\!&=\!\!& -i\,
                             \frac{g_2 g_V \,
                             e}{\sqrt{8}}\,\bar{p}_v(r^\prime)\left\{
                             - \frac{(1+\kappa_v)}{M}\left[S_\mu,S
                             \cdot\epsilon^\ast
                             \right]-\frac{1}{2 M}\,\epsilon^\ast_\mu
                             \right. \nonumber \\
& &\qquad \left.
   -\frac{1}{M \omega}\,v_\mu \left[
   \left(1+\kappa_v\right)[S\cdot\epsilon^\ast,S\cdot k] -
   \epsilon^\ast\cdot r'\right]+{\cal O}\left(\frac{1}{M^2}\right)
   \right\} n_v(r)~, \label{VVcorr} \\
\langle p|\,{\cal T} \,V\cdot\epsilon^\ast A_\mu^+\,|n\rangle^{(2)}
\!\!&=\!\!& -i\,
                             \frac{g_2 g_V
\,
                             e}{\sqrt{8}}\,\bar{p}_v(r^\prime) 
\left\{- 2\lambda\,\frac{S\cdot(r^\prime-r)}{
   (r^\prime-r)^2-m_{\pi}^2} 
\left[\frac{2\,\epsilon^\ast\cdot(l_e+l_\nu)\,
   (l_e+l_\nu)_\mu}
   {(l_e+l_\nu)^2-m_{\pi}^2}-\epsilon^\ast_\mu\right]\right. 
\nonumber
\\
\!& &+ \frac{\lambda}{M}\,\frac{\left(v\cdot r^\prime-v\cdot r\right)
   S\cdot(r+r^\prime)}{(r^\prime-r)^2-m_{\pi}^2}
   \left[\frac{2\,\epsilon^\ast\cdot(l_e+l_\nu)\,(l_e+l_\nu)_\mu}
   {(l_e+l_\nu)^2-m_{\pi}^2}-\epsilon^\ast_\mu\right]
   \nonumber \\
\!& &-2\,\lambda\left[1+\left(\frac{v\cdot l_e
+v\cdot l_\nu}{2 M}\right)\right]\frac{S\cdot
   \epsilon^\ast\,(l_e+l_\nu)_\mu}{(l_e+l_\nu)^2-m_{\pi}^2} +
\frac{\lambda}{M}\,S\cdot
   \epsilon^\ast\,v_\mu \nonumber \\
\!& &- \frac{\lambda}{M}\left[\frac{(2+\kappa_s+\kappa_v)\,
[S\cdot\epsilon^\ast,S\cdot
k]\,S^\alpha}{\omega}+\frac{(\kappa_v-\kappa_s)\,
S^\alpha\,[S\cdot\epsilon^\ast,S\cdot k]}{\omega}\right.\nonumber \\
\!& &\phantom{+\frac{\lambda}{M}} \left.\left.
   -\frac{2\,S^\alpha\epsilon^\ast\cdot r'}{\omega}\right]
   \left[g_{\mu\alpha}-\frac{(l_e+l_\nu)_\alpha(l_e +
   l_\nu)_\mu}{(l_e+l_\nu)^2-m_\pi^2}\right] 
+{\cal O}\left(\frac{1}{M^2}\right)\right\} n_v(r)~,
\label{VAcorr} 
\end{eqnarray}
with $\omega = v \cdot k$. Also, $m_\pi$ is the charged pion mass,
and $e=|e|$ is the
elementary charge.  
Turning to the SSE, we note 
that the vector--vector correlator is free of $\Delta$
effects to ${\cal O}(\epsilon^2)$, so that the leading $\Delta$(1232)
effect appears only in the vector--axial correlator, given by 
\begin{eqnarray}
\langle p|{\cal T} \,V\cdot\epsilon^\ast A_\mu^+ |n\rangle^{(2),\Delta}
\!\!\!&= \!\!\!& -i\,
      \frac{g_2 g_V\,
       e}{\sqrt{8}}\,\bar{p}_v(r^\prime) 
\left\{
- \frac{g_{\pi N\Delta}b_1}{3\, M} \right.\nonumber\\
& &\!\!
\times \left[\frac{2\Delta\,[k^\alpha
S\cdot\epsilon^\ast
   -\omega\, v^\alpha S\cdot\epsilon^\ast-\epsilon^{\ast\,\alpha}S\cdot
k]}{\Delta^2-
   \omega^2}+\frac{4\,[S\cdot\epsilon^\ast,S\cdot
k]\,S^\alpha}{3\,(\Delta-\omega)}
   \right. \nonumber \\
& & \!\! \left. \left.
    -\frac{4\,S^\alpha \,[S\cdot\epsilon^\ast,S\cdot
k]}{3\,(\Delta+\omega)}\right]
\left[g_{\mu\alpha}-\frac{(l_e+l_\nu)_\alpha(l_e+l_\nu)_\mu}{(l_e+l_\nu)^2-
   m_\pi^2}\right] +{\cal O}\left(\frac{1}{M^2}\right)\right\} n_v(r)~,
\label{VAcorrD}
\end{eqnarray}
where $g_{\pi N\Delta} = 1.05$ and $b_1 = 12.0$ 
are the leading strong and 
electromagnetic coupling constants in the coupled $N \Delta \pi \gamma$ 
system~\cite{BHM}, 
noting that $\Delta\equiv M_\Delta - M$.  
We neglect the 
radiative correction to this contribution, as the contribution itself is 
extremely small. 
One can easily check from the 
continuity equations satisfied by the correlators 
that gauge invariance is satisfied in the above equations~\cite{BHM}. 
Note that the ${(2)}$ superscript explicitly indicates that
we report the matrix elements in NLO.

\medskip

\noindent {\bf 3.} 
Let us compare our matrix elements with those of Ref.~\cite{gaponov}. 
As Eqs.~(\ref{VVcorr}-\ref{VAcorrD}) 
make apparent, only the electron bremsstrahlung contribution
makes an ${\cal O}(1)$ contribution to neutron radiative 
$\beta$--decay. Such a result is at odds with 
Ref.~\cite{gaponov} and, indeed, with the literature
on ``outer'' radiative corrections~\cite{sirlin} in neutron $\beta$--decay. 
In these papers there is an ${\cal O}(1)$ contribution from 
proton bremsstrahlung as well. The source of 
the {\it apparent} discrepancy can readily be found. 
The form of the decay amplitude 
for neutron radiative $\beta$--decay, as follows from 
computing the bremstrahlung contributions in QED~\cite{gaponov}, is
\begin{eqnarray}
{\cal M} &=& \frac{e g_V G_F i}{\sqrt{2}}
\left\{ 
\bar u_e (l_e) \frac{(2 l_e \cdot \epsilon^\ast + 
\slash{\!\!\!\!\!\; \epsilon}^\ast \slash{\!\!\!\!\!\; k} )}{2l_e \cdot k} 
\gamma_\rho(1-\gamma_5) v_\nu(l_\nu)\bar u_p(p^\prime) \gamma^\rho
(1-\lambda\gamma_5)u_n(p) \right. \nonumber \\
& & \left.-
\bar u_e (l_e) 
\gamma_\rho(1-\gamma_5) v_\nu(l_\nu)\bar u_p(p^\prime) 
\frac{(2 p^\prime \cdot \epsilon^\ast + 
\slash{\!\!\!\!\!\; \epsilon}^\ast 
\slash{\!\!\!\!\!\; k} )}{2p^\prime \cdot k} 
\gamma^\rho
(1-\lambda\gamma_5)u_n(p) \right\} \;.
\label{meqed}
\end{eqnarray}
The QED treatment 
neglects photon emission from the effective weak vertex; it 
is correct in leading order in $1/M$ only. 
Consequently, we 
consider $|{\cal M}|^2$ here in leading order only. 
Note that for each photon polarization state
$p^\prime \cdot \epsilon^\ast/p^\prime\cdot k$ is of 
${\cal O}(1/M)$, so that the proton bremsstrahlung contribution
is also of ${\cal O}(1/M)$ --- and thus negligible. 
However, in effecting the photon polarization sum, the gauge 
invariance of QED also permits the replacement 
$\sum_{\rm \sigma} \epsilon_\mu^\ast(\sigma) \epsilon_\nu(\sigma) 
\to - g_{\mu \nu}$. This suggests 
that the $p^\prime \cdot \epsilon^\prime/p^\prime\cdot k$ term,
when squared and summed over the photon helicity, yields a contribution
of ${\cal O}(1)$. This is, indeed, what happens upon explicit 
calculation. Employing lepton and hadron tensors, 
the square of the matrix element can be written as 
\begin{equation}
\sum_{\rm spins} |{\cal M}|^2 =
\frac{e^2 g_V^2 G_F^2}{2}
\left( \frac{1}{(l_e\cdot k)^2} L^{\rm ee}_{\rho \delta}H^{\rho \delta}
+ 
\frac{1}{M^\prime{}^2 \omega^2} L^{\rho \delta}H^{\rm ee}_{\rho \delta}
- 
\frac{1}{M^\prime \omega (l_e\cdot k)} M^{\rm ee,\,mixed} \right) \,,
\label{msqualo}
\end{equation}
where we have retained only the leading expression in each term. 
Employing the $g_{\mu \nu}$ replacement for the photon helicity sum,
we find 
\begin{eqnarray}
L^{\rm ee}_{\rho \delta}H^{\rho \delta} \!&=\!& -64 
M M^\prime (m_e^2 - l_e \cdot k) 
\left(
(1 + 3\lambda^2)  E_\nu ( E_e + \omega) 
+ 
(1 -\lambda^2) (\mathbf{l_e} \cdot \mathbf{l_\nu} 
+ \mathbf{l_\nu} \cdot \mathbf{k})
\right) \;, \\
L^{\rho \delta}H^{\rm ee}_{\rho \delta} \!&=\!& -64 M (M^\prime)^3
\left( (1 + 3\lambda^2) E_\nu E_e 
+ (1 -\lambda^2) \mathbf{l_\nu} \cdot \mathbf{l_e}  \right) \;,\\
M^{\rm ee,\,mixed}
\!&=\!& -64 M (M^\prime)^2
\left(  (1 +3 \lambda^2) E_\nu ( 2E_e^2 + E_e \omega - k\cdot l_e)\right. 
\nonumber \\
\!& \!& \left. \quad  + 
(1 -\lambda^2) (2E_e \mathbf{l_\nu} \cdot \mathbf{l_e} 
+ E_e \mathbf{l_\nu} \cdot \mathbf{k})
\right) \;, \label{msqualo1}
\end{eqnarray}
identical to the result of Ref.~\cite{gaponov}, save for an overall
sign. We have checked that this result is identical to that
obtained using the leading contribution from the 
$L^{\rm ee}_{\rho \delta}H^{\rho \delta}$ term exclusively, after 
explicitly summing over the photon polarization states. 
Equation (\ref{meqed})
and Eqs.~(\ref{me}-\ref{Acorr}) are 
consistent to leading order in $1/M$.
Furthermore, the leading contribution
to the outer radiative corrections in neutron $\beta$--decay 
is also from electron bremsstrahlung,  
as calculated here, complemented by the photon exchange graph ---
for a recent attempt at calculating
radiative corrections to neutron $\beta$--decay within EFT, see 
Ref.~\cite{AFNSGKM}. 

Noting the normalization of the nonrelativistic 
spinors~\cite{BHM}, 
the total decay rate is given by
\begin{equation}
\Gamma={1 \over (2 \pi)^{8}}
\int d^3\mathbf{p^\prime} \, d^3\mathbf{l_e} \, d^3\mathbf{l_\nu}\, 
d^3\mathbf{k}\, {M^\prime \over E^\prime} 
{1 \over 2 E_\nu}
{1 \over 2 E_e} {1 \over 2 E_\gamma} 
\sum_{\rm spins} 
\left|{\cal M}\right|^2 \delta ^{(4)} (p - p^\prime -l_e-l_\nu-k) \,,
\end{equation} 
or 
\begin{equation}
\Gamma={M^\prime \over 8 (2 \pi)^{8}}
\int |\mathbf{l_e}| \omega d\omega\, dE_e\, d\Omega_e 
d \Omega_k\, d \Omega_\nu
\left[
\frac{\Theta(M- E_e - E_\nu - \omega) \,E_\nu 
\sum_{\rm spins} 
\left| {\cal M}\right|^2 |_{p^\prime = p - l_e - l_\nu - k}}
{|M - E_e - \omega + \mathbf{l_e} \cdot \mathbf{n_\nu} + 
\mathbf{k} \cdot \mathbf{n_\nu}|} \right] \;,
\label{totrate}
\end{equation} 
where $\mathbf{n_\nu} \equiv 
\mathbf{\hat{l}_\nu}$ and 
\begin{equation} 
E_\nu = 
\frac{M^2 + m_e^2  - M^\prime{}^2 
- 2M(E_e + \omega) 
+ 2 E_e \omega - 2 \mathbf{l_e}\cdot \mathbf{k}}
{2(M-E_e -\omega  + \mathbf{l_e} \cdot \mathbf{n_\nu} + 
\mathbf{k} \cdot \mathbf{n_\nu})} \,.
\label{defenu}
\end{equation}
To complete the integration over the four-particle phase space, 
we let $\mathbf{\hat{l}_e}$ define the $\mathbf{z}$--direction, so that 
$\mathbf{\hat{k}}\cdot \mathbf{\hat{l}_e}\equiv x_k$ and  
$\mathbf{n_\nu}\cdot \mathbf{\hat{l}_e}\equiv x_\nu$. Thus
Eq.~(\ref{totrate}) can be cast in the form 
\begin{eqnarray}\label{gamma}
\Gamma (\omega^{\rm min})
&=& \frac{M^\prime}{4(2\pi)^6} 
\int_{\omega^{\rm min}}^{\omega^{\rm max}} \omega d\omega
\int_{m_e}^{E_e^{\rm max}(\omega)} |\mathbf{l_e}| dE_e 
\int_{x_k^{\rm min}\!(E_e,\omega)}^{x_k^{\rm max}\!(E_e,\omega)}
dx_k 
\int_{-1}^{1} d x_\nu \int_0^{2\pi} d\phi_- \nonumber
\\ && \qquad\qquad \times
\frac{E_\nu}{| M - E_e - \omega + |\mathbf{l_e}| x_\nu + 
\mathbf{k} \cdot \mathbf{n_\nu}|}
%{1 \over 2} 
\sum_{\rm spins} \left|
{\cal M}\right|^2 |_{p^\prime = p - l_e - l_\nu - k} \;,
\end{eqnarray}
where $\phi_-\equiv \phi_k - \phi_\nu$. The lowest photon energy, 
$\omega^{\rm min}$, is determined by the energy resolution of
the detector; thus the total decay rate depends on 
$\omega^{\rm min}$. We have 
\begin{equation}
\omega^{\rm max} = \frac{(M-m_e)^2 - M^\prime{}^2}{2(M - m_e)} \quad ; \quad
E_e^{\rm max}(\omega) = \frac{M^2 + m_e^2 - 
M^\prime{}^2 - 2M\omega}{2(M - \omega(1 + \beta_e))}
\,.
\end{equation}
The $\beta_e$ dependence in $E_e^{\rm max}$, noting 
$\beta_e \equiv |\mathbf{p_e}|/E_e$, implies that $E_e^{\rm max}$ is  
determined numerically, by iterating to a 
self-consistent solution for fixed $\omega$. 
The range in $x_k$ is determined by demanding that $E_\nu \ge 0$, 
i.e., by requiring
\begin{equation}
(M^2 + m_e^2 - M^\prime{}^2)\frac{1}{2} + E_e \omega - M(E_e + \omega)
-\mathbf{l_e}\cdot\mathbf{k} \ge 0 \,,
\end{equation}
as well as by demanding that $M - M^\prime - E_e - E_\nu - \omega \ge 0$. 

We also compute the polarization of the emitted photon. 
Defining the polarization states
$\epsilon_{1}^\mu=(0,-\sin\phi_k,\cos \phi_k,0)$ and 
$\epsilon_{2}^\mu=(0,\cos\theta_k\cos \phi_k, \cos\theta_k\sin\phi_k,
-\sin\theta_k)$, 
we can, in turn, define states of 
circular polarization, namely, 
$\epsilon_L \equiv (\epsilon_{1} + i \epsilon_{2})/\sqrt{2}$ and 
$\epsilon_R \equiv (\epsilon_{1} - i \epsilon_{2})/\sqrt{2}$. 
With these conventions, $\epsilon_L$, e.g., does indeed correspond
to a left-handed photon when $\mathbf{k} \parallel \mathbf{l_e}$. 
We define the polarization $P$ via 
\begin{equation}
P = \frac{\Gamma_R - \Gamma_L}{\Gamma_R + \Gamma_L}\;. 
\end{equation}
We can also study the polarization
as a function of $\omega$ and $E_e$ as well; in such cases, we 
define $P(\omega)$ by replacing $\Gamma_{L,R}$ with 
$d \Gamma_{L,R}/d\omega$ and 
$P(\omega,E_e)$ by replacing $\Gamma_{L,R}$ with 
$d^2 \Gamma_{L,R}/d\omega dE_e$. 

\medskip

\begin{figure}[ht]
\begin{center}
\includegraphics[width=3.5in]{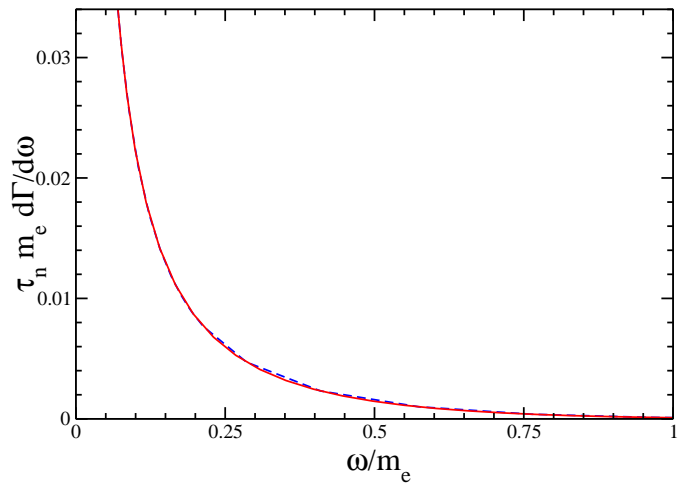}
\vspace{-0.2cm}
\caption{The photon energy spectrum for radiative neutron $\beta$--decay. 
The dashed line denotes the result 
to NLO in the SSE, 
whereas the solid line denotes the leading order result, 
determined using Eqs.~(\ref{msqualo}-\ref{msqualo1}), 
employed in 
Ref.~\cite{gaponov}.
\label{fig:spec}}
\end{center}
\end{figure}

\noindent {\bf 4.} 
We can now present our results. For definiteness, we
specify the input parameters. We use~\cite{pdg2002,BHM}: 
$G_F = 1.16639\cdot 10^{-5} \hbox{GeV}^{-2}$, $\alpha^{-1} = 137.03599976$, 
noting $\alpha=e^2/(4\pi\hbar c)$ in the Heaviside-Lorentz convention, 
$m_e = 0.510998902\, \hbox{MeV}$,  $m_\pi = 139.57018\,  \hbox{MeV}$,
$M = 939.56533\, \hbox{MeV}$, $M^\prime =938.27200\, \hbox{MeV}$,
$V_{ud} = 0.9740$, $\Delta_R^V = 0.0240$~\cite{towner}, $\lambda= 1.267$, 
$\kappa_v = 3.706$, $\kappa_s =- 0.120$, 
$F_\pi= 92.3\, \hbox{MeV}$, $g_{\pi NN}=13.10$, 
$r_A= 3.395\cdot 10^{-3}\, \hbox{MeV}^{-1}$, 
$M_\Delta = 1232\, \hbox{MeV}$, $g_{\pi N \Delta}=1.05$, $b_1 =12.0$, 
and the neutron lifetime $\tau_n = 885.7\, \hbox{s}$.
We show the photon energy 
spectrum $d\Gamma / d\omega$ in Fig.~\ref{fig:spec}, 
and for the total branching ratio, which depends 
on the range chosen for $\omega$, we find, 
\begin{eqnarray}
\omega \in [0.005\, \hbox{MeV}, 0.035\, \hbox{MeV}]\,, &\quad& 
\hbox{Br}: 5.17 \cdot 10^{-3} \,, \nonumber 
\\
\omega \in [0.035\, \hbox{MeV}, 0.100\, \hbox{MeV}]\,, &\quad& 
\hbox{Br}: 2.21 \cdot 10^{-3}\,, \\
\omega \in [0.100\, \hbox{MeV}, \omega^{\rm max}=0.782\, \hbox{MeV}]\,, &\quad&
\hbox{Br}: 1.44 \cdot 10^{-3} \,.
\nonumber
\end{eqnarray}
The branching ratio determined for 
$\omega \in [0.035\, \hbox{MeV}, 0.100\, \hbox{MeV}]$ can be compared
directly with the experimental limit of 
${\rm Br} < 6.9 \cdot 10^{-3}\, (90\% \hbox{CL})$~\cite{mbeck},
with which it is compatible. However, the branching ratio 
for this range of $\omega$, as well as the 
photon energy spectrum for $\omega/m_e$ greater than $\simeq 0.2$, 
are roughly a factor of two larger 
than the numerical results reported in Ref.~\cite{gaponov}. 
The discrepancy appears to grow smaller as the photon energy
goes to zero. Note, too, that we retain the complete expression
for $\sum_{\rm spins}|{\cal M}|^2$ 
in our subsequent numerical calculation; Ref.~\cite{gaponov} 
approximates the integration over phase space and retains the
term proportional to $1+3\lambda^2$ only. 
Note that the approximate angular integrals
in Eq.~(19) in the
first paper of Ref.~\cite{gaponov}
are correct only if $E^\prime$ (in our notation)
is replaced by $M^\prime$, as they neglect $|\mathbf{p^\prime}|$ 
relative to $E^\prime$. 
However, the authors then proceed to integrate over $E^\prime$ in Eq.~(20), 
which is incompatible with the approximation of Eq.~(19). 
We emphasize that the discrepancy 
is not due to the recoil-order 
corrections --- in Fig.~\ref{fig:spec} we superimpose the numerical
results we find using the leading order form of 
$\sum_{\rm spins} |{\cal M}|^2$, given in Eqs.~(\ref{msqualo}-\ref{msqualo1}). 
The two curves can scarcely be distinguished; indeed, the recoil-order
corrections are no larger than ${\cal O}(0.5\%)$. The SSE contribution
is itself of ${\cal O}(0.1\%)$. In constrast, the recoil-order 
corrections to the $A$ and $a$ correlations in neutron $\beta$--decay 
are of ${\cal O}(1-2\%)$~\cite{chisvg}; apparently, 
the appearance of an additional particle in the final state 
makes the recoil-order corrections, which are controlled by
the dimensionless parameter $\epsilon$, smaller still. 

\vspace{0.5cm}

\vspace{0.5cm}
\begin{figure}[ht]
\begin{center}
\includegraphics[width=3.5in]{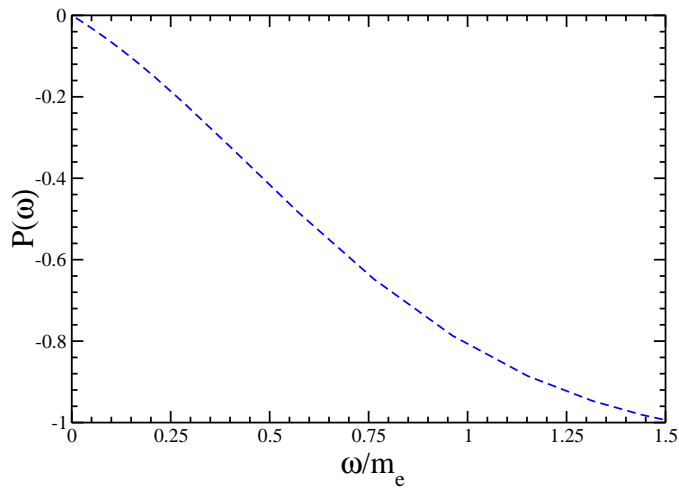}
\vspace{-0.2cm}
\caption{The photon polarization $P(\omega)$ 
in radiative neutron $\beta$--decay to NLO in the SSE.
\label{fig:asym}}
\end{center}
\end{figure}

\begin{figure}[tb]
\begin{center}
\includegraphics[height=2.5in]{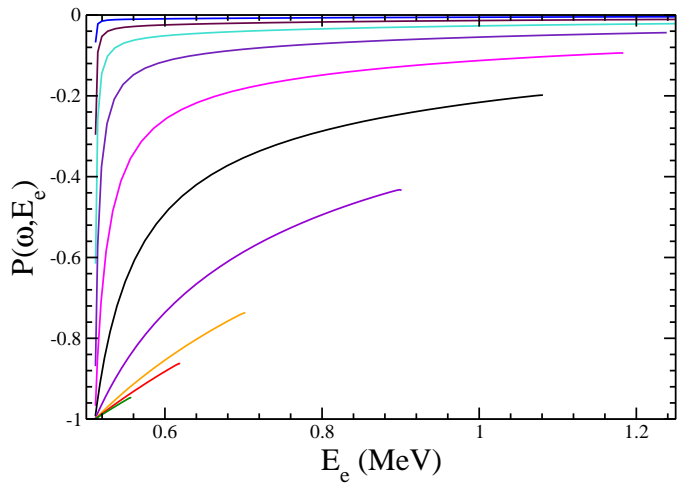}
\vspace{-0.2cm}
\caption{The photon polarization $P(\omega,E_e)$ 
in radiative neutron $\beta$--decay 
to NLO in the SSE,
as a function
of $E_e$ for $(E_e^{\rm max}- E_e)/E_e^{\rm max} \gtrsim 0.2\%$ and 
various, fixed $\omega$. For $E_e$ such 
that $(E_e^{\rm max}- E_e)/E_e^{\rm max} \lesssim 0.2\%$, 
the polarization plunges to $-1$, see text. The curves from smallest
absolute polarization to largest have 
$\omega = 0.00539, 0.0135, 0.0265, 0.0534, 
0.109, 0.209, 0.390, 0.589, 0.673$, and $0.736$ MeV, respectively. 
\label{fig:asymee}}
\end{center}
\end{figure}

We present the photon polarization in Fig.~\ref{fig:asym}. The
polarization evolves from near-zero at low photon energies to 
nearly 100\% left-handed polarization at high photon energies,
as consistent with the discussion of Ref.~\cite{marglau}. 
The evolution of the polarization with $\omega$ 
is dissected in Fig.~\ref{fig:asymee}; as $\omega$ grows large, 
the associated electron momentum is pushed towards zero, and the
absolute polarization grows larger. This follows as in the
circular basis we can replace 
$(2\epsilon_{\pm}^\ast\cdot l_e - {\slash \!\!\!\!\!\; k} \,
{\slash \!\!\!\!\!\; \epsilon_\pm^\ast})$ in 
$\langle \bar \nu_e e^- \gamma | J_\mu^- | 0 \rangle$ of 
Eq.~(\ref{lepcurre})
with $(2\epsilon_{\pm}^\ast\cdot l_e - \omega ( 1\pm \gamma_5)\gamma^0 
{\slash \!\!\!\!\!\; \epsilon_\pm^\ast})$ with 
$\epsilon_{+,-}=\epsilon_{R,L}$. 
The photon associated with the first term has no
circular polarization; this contribution vanishes if $|\mathbf{l_e}|=0$. 
In this observable as well
the ${\cal O}(1/M)$ contributions are ${\cal O}(0.5\%)$ or less.
Interestingly, the inclusion of these contributions does not 
impact the determined polarization to an appreciable 
degree when $\mathbf{l_e} \parallel \pm \mathbf{k}$; 
$P\approx -1$. Note 
that as 
$E$ approaches $E_e^{\rm max}(\omega)$, 
$\mathbf{l_e}$ becomes parallel to $- \mathbf{k}$, so that 
$\epsilon^\ast \cdot l_e$ approaches zero and $P$ approaches
$-1$ to a high degree of accuracy. 
In neutron radiative $\beta$--decay, the polarization can differ
appreciably from unity, so that the calculation of the 
polarization is {\it necessary} 
to realize a SM test; significant deviations from this prediction would
nevertheless signify the palpable presence 
of a left-handed anti-neutrino or of non-$V-A$ currents. As noted
by Martin and Glauber~\cite{marglau}, the polarization of the photon
in $S$-state orbital electron capture is also sensitive
to the {\it phase} of the vector and axial-vector couplings
in the low-energy interaction Hamiltonian~\cite{leeyang} if the
anti-neutrino is no longer assumed to be strictly right-handed. 
Such expectations apply to neutron radiative $\beta$--decay as well, 
so that the photon polarization can probe new physics effects 
to which the correlation coefficients in neutron $\beta$--decay
are insensitive~\cite{jackson}. 
\vspace{0.5cm}

\medskip

\noindent {\bf 5.} 
In this letter, we have computed the photon energy spectrum
and photon polarization in neutron radiative $\beta$--decay in 
an effective field theory approach, utilizing 
heavy baryon chiral perturbation theory and the small-scale
expansion, including all terms in ${\cal O}(1/M)$. 
The leading contribution to the photon energy spectrum has
been calculated previously~\cite{gaponov}; we agree with the expression
in Ref.~\cite{gaponov} 
for $\sum_{\rm spins}|{\cal M}|^2$, though we disagree with
their numerical results for the photon energy spectrum. 
Moreover, we find that the $O(1/M)$ terms are numerically quite small, 
generating contributions no larger than ${\cal O}(0.5\%)$, so 
that radiative neutron $\beta$--decay is quite insensitive to 
nucleon structure effects beyond those encoded in $g_V$ and $g_A$ 
--- and offers no clear resolution of the muon radiative capture problem. 
On the other hand, we have found that nucleon structure effects have
a similarly negligible role in the determination of the photon polarization,
so that a precise measurement of the 
photon polarization may well offer a crisp diagnostic 
of non-SM effects. Such studies may complement other new physics searches. 
For example, the (pseudo-T-odd) transverse muon polarization $P_\mu^\perp$ in 
$K^+\to \mu^+ \nu \gamma$ decay is sensitive to large 
squark generational mixings in
generic supersymmetric models~\cite{ng} --- such charged-current processes
survive flavor-changing-neutral-current (FCNC) 
bounds~\cite{ng}. 
The mechanisms discussed in Ref.~\cite{ng} modify the photon polarization 
as well, and can also act to modify the $d\to u$ charged, weak current at
low energies, to impact the photon polarization, as is our concern here. 
Finally, we note that 
the polarization  of the photon 
in radiative B-meson decay, namely in $b\to s \gamma$ decay, is also 
left-handed in the SM, modulo ${\cal O}(1/M_B)$ corrections, estimated
to be of order of a few percent; 
it is also sensitive to non-SM operators~\cite{pirjol}, 
as we have discussed here.

\vskip 1cm

\noindent{\large {\bf Acknowledgements}}

\smallskip\noindent
The work of S.G. is supported in
part by the U.S. Department of Energy under contract number 
DE-FG02-96ER40989. We are grateful to Fred Myhrer for providing us
with a draft of Ref.~\cite{AFNSGKM} prior to publication and thank
Gudrun Hiller for a discussion of the work in Ref.~\cite{ng}. 
We also thank the members of the Fundamental Neutron Physics Program 
in the Ionizing Radiation Division of the 
National Institute of Science and Technology for a discussion of
the experimental possibilities.

%%%%%%%%%%%%%%%%%% REFERENCES %%%%%%%%%%%%%%%%%%%%%%%%%%%%
%%\newpage

\end{document}